\documentclass[11pt]{article}
\usepackage{epsfig}
\usepackage{epic}
\newcommand{\be}{\begin{equation}}
\newcommand{\ee}{\end{equation}}
\newcommand{\bea}{\begin{eqnarray}}
\newcommand{\eea}{\end{eqnarray}}

\newcommand{\vs}[1]{\vspace{#1 mm}}
\newcommand{\hs}[1]{\hspace{#1 mm}}

\begin{document}
\thispagestyle{empty}

\rightline{hep-th/0303232}

\vs{20}

\centerline{\large\bf Non-Standard Intersections of S-Branes in 
$D=11$ Supergravity} 
\vs{15}
\centerline{Nihat Sadik Deger\footnote{e-mail: deger@gursey.gov.tr}}
\vs{7}
\centerline{\sl Feza Gursey Institute} 
\vs{1}
\centerline{\sl Emek Mah. No:68, Cengelkoy}
\vs{1}
\centerline{\sl 81220 Istanbul-Turkey}
\vs{25}

\begin{abstract}
We construct new intersecting S-brane solutions in
11-dimensional supergravity which do not have 
supersymmetric analogs. They are obtained by letting
brane charges to be proportional to each other. 
Solutions fall into two categories with respect to whether there is a 
non-diagonal term to be cancelled in the field equations or not. In each 
case we show that they can be constructed by using a simple set 
of rules which is similar to the harmonic function rule of the usual 
static 
p-branes. Furthermore, we study an intersection where the Chern-Simons 
term 
makes a non-zero contribution to the field equations. We show that this 
configuration has a singularity like other S-branes. 
\end{abstract} 

\newpage

\setcounter{page}{1}

\section{Introduction}

S-branes are a particular class of D-branes \cite{polc} where along the 
time-like direction Dirichlet boundary condition 
is applied \cite{s1}. Therefore they
exist only for a moment in time. Recently effort to understand these 
objects increased considerably (see, e.g. 
\cite{s2}-\cite{rec9}) since they are 
interesting both as time-dependent backgrounds and for tachyon 
condensation. They may also be useful in establishing 
the $dS/CFT$ duality \cite{strominger}. 

\
\

Although the name $S$-$brane$ first appeared in \cite{s1} and more 
examples were obtained subsequently in \cite{s2, s3}, such 
supergravity solutions were studied before \cite{behrndt, pope1, 
pope2,ovrut1, ovrut2} where the primary motivation was cosmology. 
Additionally, let us note that similar solutions 
were also investigated by relating them to Toda-like systems 
\cite{toda2, toda3}. (See \cite{toda4} for a review and more references.)

\
\

After the construction of single S-brane solutions \cite{s1, s2, s3} a 
natural next step was 
to 
consider their intersections.
In \cite{s3}, multiply charged SD$p$/SD($p-2$) and
SD$p$/SD($p-4$) brane solutions were constructed.
In \cite{deger} all possible orthogonally
intersecting S-brane solutions of $D=11$ supergravity theory
corresponding to supersymmetric M-brane intersections were obtained. 
Later in \cite{ohta}
the S-brane intersection rules were generalized to arbitrary dimensions.
In \cite{deger,ohta} it was observed that all solutions can be obtained by 
a simple
procedure of multiplying the brane and the transverse directions by
certain powers of two hyperbolic functions of time. This is the S-brane 
analog of the 
harmonic function rule of the usual p-branes \cite{har1, har2, har3, 
har4}.

\
\

The S-brane intersections that are obtained in \cite{deger} are in 1-1 
correspondence with the supersymmetric intersections of M-branes. (For 
instance, there is $SM2\perp SM5(1)$ instead of $M2\perp M5(1)$.)
In this paper we will call those solutions $standard$ and show that 
actually 
there exist more intersections without supersymmetric duals.
Since, S-branes anyway break all supersymmetry, there is a 
priori 
no reason for neglecting these cases.  

\
\

In standard intersections in $D=11$ each brane  
has 3 free parameters and the intersection rules derived in 
\cite{deger} 
are found by assuming that all parameters are independent from 
each other. 
However, we will show in this paper that relaxing this 
condition allows other intersections as well. 
\footnote {This possibility was mentioned in \cite{ohta} and the solutions 
that we will construct in section 2.1  
correspond 
to block-orthogonal intersections in the Toda approach \cite{toda3}.}
In particular, if brane charges are proportional to each other, then 
more solutions are possible. 

\
\
 
The organization of this paper is as follows. In the next section we will 
begin construction of 
non-standard intersections in which the Chern-Simons term doesn't effect 
the 
field equations. 
They  
fall into two 
categories with respect to 
whether there is a
non-diagonal term to be cancelled in the field equations or not. 
In the second one we only have a condition on the magnitudes of the
charges, whereas in the first we also need oppositely charged branes in
the solution.
In each
subclass we find that solutions can be constructed by using a simple set
of rules which is similar to the set of rules found for standard 
intersections 
\cite{deger} upto 
some small modifications. In section 3 we will
analyze an example, $SM2\perp SM5\perp SM2(-1)$, where the Chern-Simons 
term makes 
a non-trivial contribution to 
the field equations and imposes some 
constraints unlike any other solution studied in the literature so far.
The 
equations are quite hard to solve analytically, however using numerical 
methods 
we will show that there is still singularity. We will conclude in section 
4 with some comments and possible future directions.

\
\

\section{Intersections with $F\wedge F=0$}

\
\

The bosonic action of the 11-dimensional supergravity can be written as
\be
S=\int d^{11}x(\sqrt{-g}R-\frac{1}{2.4!}\sqrt{-g} F^2 - \frac{1}{6}
F\wedge F \wedge A)
\ee
where the last term is the Chern-Simons term.
The equations of motion are 
given by
\bea\label{eins1}
R_{AB}&=&\frac{1}{2.3!}F_{ACDE}F_{B}{}^{CDE}-\frac{1}{6.4!}g_{AB}F^{2},\\
d*F&=&\frac{1}{2}F\wedge F.
\label{eins2}
\eea

\
\

We consider a metric of the following form
\be
ds^2=-e^{2A}dt^2\,+\,\sum_k\,e^{2C_k}\,ds_k^2\,+\,e^{2D}\,d\Sigma_{n,\sigma}^2
\label{metric}
\ee
where $ds_k^2$ (for $k=1,2..$) is the metric on the $d_k$ dimensional flat
space. $d\Sigma_{n,\sigma}^2$ is the metric of the
$n$-dimensional unit sphere $\sigma=1$, unit hyperbola $\sigma=-1$ or
flat space $\sigma=0$. The metric functions depend only on $t$.

\
\

In this section we are interested in solutions
where $F\wedge F=0$ and thus the 4-form field equations reduce to 
$dF=d*F=0$. The spatial transverse space to
any SM2-brane is 7-dimensional and if we denote the
closed
volume-form of this space with $\textrm{Vol}_7$,  
then the 4-form field of the $i$'th SM2-brane can be
written as  
\be\label{fm2}
F_i=q_i\,\,*\,\textrm{Vol}_7,
\ee
where $*$ is the Hodge dual with respect to the full metric. Similarly,
for each SM5-brane, the space-like transverse space is
4-dimensional. Defining $\textrm{Vol}_4$ to be the volume-form of this 
space,
the 4-form field
corresponding to the j'th SM5-brane is equal to
\be\label{fm5}
F_j=q_j\,\,\textrm{Vol}_4.
\ee
It is easy to see that we have $dF=0$ and $d*F=0$ for both cases. 

\
\

One can
simplify the Ricci tensor by fixing $t$-reparametrization invariance so
that

\be
A=\sum_{k}d_{k}C_{k}\,+\,nD.
\label{gauge}
\ee

With respect to the
orthonormal frame $E^{0}=e^{A}dt$, $E^{\alpha_k}=e^{C_k}dy^{\alpha_k}$
and $E^{\theta}=e^{D}e^{\theta}$, where $e^{\theta}$ is an orthonormal
frame on $\Sigma_{n,\sigma}$ and with the above gauge choice the Ricci 
tensor becomes \cite{deger}

\bea
R_{00} & = & e^{-2A}\left[-A^{''}+A^{'2}-\sum_{k}d_{k}C_{k}^{'2}-nD^{'2}\right]\nonumber\\
R_{\alpha_k\beta_k} & = & e^{-2A}
\left[C_{k}^{''}\right]\delta_{\alpha_k\beta_k}\,\,\,k=1,2...\\
R_{\theta_1\theta_2} & = & e^{-2A}\left[D^{''}\right]
\delta_{\theta_1\theta_2}\,+\,\sigma(n-1)e^{-2D}
\delta_{\theta_1\theta_2},\nonumber
\label{r}
\eea
where all derivatives are with respect to the time coordinate $t$.

\
\

The curvature scalar is
\be
R=e^{-2A}\left[2A^{''}-A^{'2}+\sum_kd_kC_k^{'2}\,+\,n\,D^{'2}\right]\,+\sigma
n(n-1)\,e^{-2D}.
\label{curvature}
\ee

\
\

Now we are ready to ask which configurations are allowed? There are 3 
restrictions 
coming from our choice of the metric (\ref{metric}) and the field 
equations (\ref{eins1}) and (\ref{eins2}):

\
\

(i) The dimension of the transverse space without time should be at least 
two, 
i.e., $n\ge2$. This is necessary in order to be able to solve the (00) 
component of the Ricci tensor. This rules out $SM5 \perp SM5(0)$, $SM5 
\perp SM5(1)$, $SM2 \perp SM5(2)$ and $SM2 \perp SM5(-1)$. 

\
\

(ii) In this section we assume that 
$F\wedge F =0$. This eliminates single $SM2 \perp SM5(2)$ pair.

\
\

(iii) Our Ricci tensor is diagonal and therefore 
there shouldn't be any term at the right hand side of the Einstein's 
equation (\ref{eins1})
coming from the $F_{AMNP}F_{B}{}^{MNP}$  contraction. This excludes 
single $SM5 \perp SM5(4)$ and single $SM2 \perp SM2(1)$ pairs. 

\

Taking into account these restrictions we are left with: 
\bea \label{standard}
\textrm{Standard Pairs} &:& SM2\perp SM2(0), SM2\perp 
SM5(1),SM5\perp SM5(3) \\
\textrm{Non-Standard Pairs} &:& SM2\perp SM2(-1), SM2\perp SM5(0)
\label{non}
\eea
By $standard$ we mean intersections which have supersymmetric analogs. 
All possible 
cases when each brane is making a standard intersection with others, have 
been constructed in \cite{deger}. In the next subsection we will construct 
all allowed combinations between the two groups listed above.
The constraints (ii) and (iii) can be surmounted by adding more pairs 
which 
contributes in the same non-diagonal direction but with
opposite sign. We leave the investigation of these to the subsection 2.2.

\subsection{Intersections without any non-diagonal terms}

\
\

In this part we will first construct $SM2\perp SM2(-1)$ 
and $SM2\perp SM5(0)$ solutions  and
then give rules to construct intersections containing more branes.
Formulas will be given in the most general form whenever possible.
For $SM2\perp SM2(-1)$ we have n=4 and $(-1)$ means that they don't have 
any 
common tangent direction. (They share only the moment of their existence.)
Let the first 
SM2-brane be located at $(x_1,x_2,x_3)$ with charge $q_1$ 
and another SM2-brane be located at $(x_4,x_5,x_6)$
with charge $q_2$. Then, the metric (\ref{metric}) becomes,
\be 
ds^2=-e^{2A}dt^2\,+\, 
e^{2C_1}\,(dx_1^2+dx_2^2+dx_3^2)+\,e^{2C_2}\,(dx_4^2+dx_5^2+dx_6^2)
+\,e^{2D}\,d\Sigma_{4,\sigma}^2
\label{metric2}
\ee

We have already explained above how to solve the 4-form field equation 
(\ref{eins2}). From the 
spatial components of Einstein's equations (\ref{r}) we get:

\bea
\label{-1}
C_1''&=&-{q_1^2\over3} e^{-6C_2-8D+2A} + {q_2^2\over6} e^{-6C_1-8D+2A}\\
\label{-2}
C_2''&=&{q_1^2\over6} e^{-6C_2-8D+2A} - {q_2^2\over3} e^{-6C_1-8D+2A}\\
\label{-3}
D''+3\sigma e^{-2D+2A}&=& {q_1^2\over6} e^{-6C_2-8D+2A} + {q_2^2\over6} 
e^{-6C_1-8D+2A}
\eea

Let us define function $g$ as: 

\be
(n-1)g=A-D
\label{g}
\ee
Now, we see that this system of equations 
is solvable if we choose $C_1=C_2=-{f\over 3}$, $D=g+{2\over 3}f$ and 
$A=4g+{2\over 3}f$ together with $q_1^2=q_2^2=2Q^2$. Note that this 
is consistent with our gauge condition (\ref{gauge}). Then, 
\bea \label{reduce1}
f''&=& Q^2 e^{-2f}\\
g''&=&-\sigma (n-1) e^{2(n-1)g}
\label{reduce2}
\eea
These two equations have integrability conditions,
\bea
(f')^2+Q^2 e^{-2f}=M_1^2\\
(g')^2+\sigma e^{2(n-1)g}= M_2^2
\label{g2}
\eea
The relation between integration constants $(M_1,M_2)$ is found from the 
$R_{00}$ 
component of the Ricci tensor to be :
\be
M_1^2 \sum_{k} q_k^2 =2n(n-1) M_2^2 Q^2
\label{integrability}
\ee
If we define $H=e^{2f}$ and $G_{n,\sigma}=e^{-2(n-1)g}$, from equations 
(\ref{reduce1}) and (\ref{reduce2}) 
we get
\be\label{H}
H=\frac{Q^2}{M_1^2}\,\cosh^2\,\left[\,M_1\,(t-t_0)\,\right].
\ee
and
\be\label{G}
G_{n,\sigma}=\begin{cases}
{
M_2^2\sinh^2\left[M_2(n-1)\,\,t\right]\hs{5}
\sigma=-1 \,\,\,\,\textrm{(hyperbola)},\cr\cr
M_2^2\cosh^2\left[M_2(n-1)\,\,t\right] 
\hs{5}
\sigma=1\,\,\,\,\textrm{(sphere)},\cr\cr
e^{2M_2(n-1)\,\,t} \hs{31}\sigma=0 
\,\,\,\,\textrm{(flat)},
}
\end{cases}
\ee

\
\
\

where the translational invariance in time is used to remove one constant 
in 
(\ref{G}).
The final form of the $SM2\perp SM2(-1)$ solution is 
\bea
ds^2&=& \label{hodge}H_1^{-1/3}H_2^{1/6}(dx_1^2+dx_2^2+dx_3^2)\, +
H_1^{1/6}H_2^{-1/3}(dx_4^2+dx_5^2+dx_6^2) \nonumber \\
&+&H_1^{1/6}H_2^{1/6}G_{4,\sigma}^{-4/3}
(\,-dt^2\,+\,G_{4,\sigma}\,d\Sigma_{4,\sigma}^2\,)  \\
F&=&q_1*(dx_4\,dx_5\,dx_6\Omega_4)\,\,+\,\,q_2*(dx_1\,dx_2\,dx_3\Omega_4).
\eea
Here $d\Sigma_{4,\sigma}^2$ is the metric on the unit sphere,
the unit hyperbola or flat space and $\Omega_4$ is its volume-form.  
The hodge dual is with respect to the full metric (\ref{hodge}).
Let us remind that for this intersection $n=4$ and we have the 
conditions $H_1=H_2=H^2$ and $q_1^2=q_2^2=2Q^2$, which implies 
$M_1^2=6M_2^2$. The purpose of writing the solution as above with $H_1$ 
and $H_2$ is that 
the same form appears in all other intersections.

\
\

The construction of $SM2 \perp SM5(0)$ follows the same steps. Let $q_1$
and $q_2$ be the charges of the SM2 and SM5-branes respectively. Then
the solution can be written as
\bea
ds^2&=& H_1^{-1/3}H_2^{-1/6}\left[dx_1^2 \right]\,+\,
H_1^{-1/3}H_2^{1/3}\left[dx_2^2 + dx_3^2\right]\,+\,H_{1}^{1/6}H_{2}^{-1/6}
\left[dx_4^2+..+dx_8^2\right]\nonumber\\
&+&H_1^{1/6}H_2^{1/3}\,G_{2,\sigma}^{-2}\left[-dt^2\,+\,G_{2,\sigma}\,d
\Sigma_{2,\sigma}^2\right],\\ 
F&=&q_1 *(dx_4...dx_8\,\Omega_2)\,\,+\,\,q_2 (dx_2\,dx_3\,\Omega_2).
\eea
The SM2-brane is oriented along $(x_1,x_2,x_3)$, and SM5-brane is
oriented along $(x_1,x_4,..,x_8)$ hyperplanes. Again we have 
$H_1=H_2=H^2$ and $q_1^2=q_2^2=2Q^2$, where $H$ and $G$ are defined in 
(\ref{H}) and (\ref{G}) with $n=2$. This solution was also given in 
\cite{toda3}.

\
\

It is straightforward to generalize this method to obtain other 
intersections. 
By assuming that each term at the right-hand side of the Ricci tensor 
(\ref{r}) has
the same exponential, one gets ratios between charges. Then, choosing
all $C_k$'s to be a suitable multiple of a function $f$ (some of them
might be zero)  and using (\ref{g}),
all spatial components of Einstein equations are reduced to the 
equations (\ref{reduce1}) and (\ref{reduce2}). From the time 
component of the Ricci tensor the integrability condition 
(\ref{integrability}) is obtained.

\
\

The maximum number of SM5 and SM2-branes one can use is 4 and 8 
respectively. For their combinations the maximum numbers are
(1 SM5+ 6 SM2), (2 SM5 + 4 SM2) and (3 SM5+ 3 SM2).
There are at least 18 non-standard intersections which involve only SM2's, 
2 intersections with only SM5's and 30 intersections with both.   
We constructed all intersections which have upto 5 branes and couple of 
typical ones in the 5'th order.
After doing this we found that all of them follow the same 
pattern and can be constructed by using a simple set of rules (we 
expect these to be valid in the remaining ones too) which are:

\
\

(i) Each additional brane to the system should make 
either standard (\ref{standard}) or non-standard (\ref{non}) intersection 
with every other brane. 

\
\

(ii) Always $n$, the dimension of the transverse space without time, has 
to satisfy $n\geq2$. 

\
\

(iii) For charges, we have two distinct cases. If a system contains an 
SM2-brane which makes more than one non-standard intersection, 
\footnote{In solutions of 
this subsection there 
can be at most one brane making two non-standard
intersections and it has to be an SM2-brane, e.g. $(x_1, x_2, x_3)$ with 
$(x_4, x_5, x_6), (x_1, x_7, x_8)$. One can add (1 SM5+2 SM2) or 4 SM2's 
to 
this system.} then the charge 
of that brane is $4Q^2$. In this case, the charges of branes making only 
one 
non-standard intersection is $3Q^2$. Otherwise,  
each brane making one non-standard 
intersection carries charge $2Q^2$ (figure 1). In both cases a brane 
making only 
standard intersections has charge $Q^2$. For 
example, in the solution where SM2-branes are located at $(x_1,x_2,x_3)$, 
$(x_4,x_5,x_6)$
and $(x_4,x_7,x_8)$ with charges ($q_1$, $q_2$, $q_3$) respectively, we 
have
$q_1^2=4Q^2$, $q_2^2=q_3^2=3Q^2$. Whereas, in another triple 
intersection of SM2's with locations $(x_1,x_2,x_3)$, $(x_4,x_6,x_7)$ and 
$(x_1,x_4,x_5)$ charges are $q_1^2=q_2^2=2Q^2$ and $q_3^2=Q^2$.

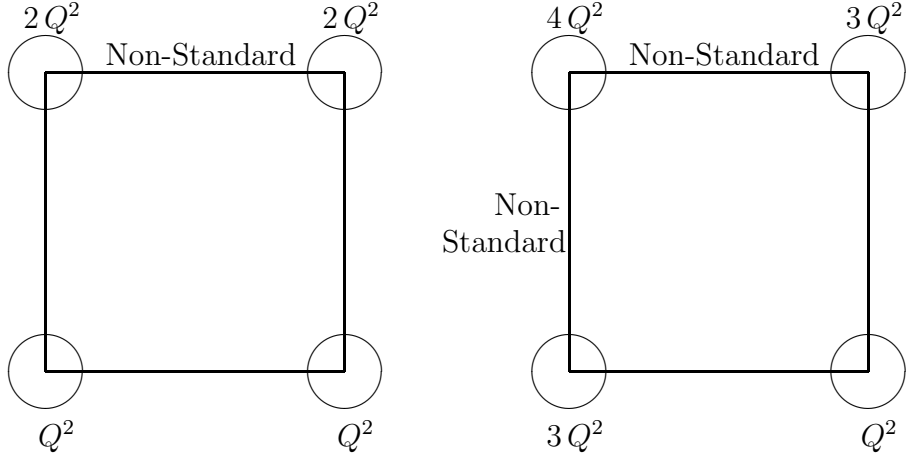
\begin{figure}
\begin{center}
\setlength{\unitlength}{0.00087489in}
\begingroup\makeatletter\ifx\SetFigFont\undefined%
\gdef\SetFigFont#1#2#3#4#5{%
  \reset@font\fontsize{#1}{#2pt}%
  \fontfamily{#3}\fontseries{#4}\fontshape{#5}%
  \selectfont}%
\fi\endgroup%
{\renewcommand{\dashlinestretch}{30}
\begin{picture}(5509,2700)(0,-10)
\put(233,450){\circle{450}}
\put(2033,450){\circle{450}}
\put(3383,450){\circle{450}}
\put(5183,450){\circle{450}}
\put(3383,2250){\circle{450}}
\put(233,2250){\circle{450}}
\put(2033,2250){\circle{450}}
\put(5183,2250){\circle{450}}
\drawline(3383,450)(5183,450)(5183,2250)
        (3383,2250)(3383,450)
\drawline(233,2250)(2033,2250)(2033,450)
        (233,450)(233,2250)
\put(2618,1170){\makebox(0,0)[lb]{\smash{{{\SetFigFont{12}{14.4}{\rmdefault}{\mddefault}{\updefault}Standard}}}}}
\put(2933,1395){\makebox(0,0)[lb]{\smash{{{\SetFigFont{12}{14.4}{\rmdefault}{\mddefault}{\updefault}Non-}}}}}
\put(593,2295){\makebox(0,0)[lb]{\smash{{{\SetFigFont{12}{14.4}{\rmdefault}{\mddefault}{\updefault}Non-Standard}}}}}
\put(3743,2295){\makebox(0,0)[lb]{\smash{{{\SetFigFont{12}{14.4}{\rmdefault}{\mddefault}{\updefault}Non-Standard}}}}}
\put(5048,2520){\makebox(0,0)[lb]{\smash{{{\SetFigFont{12}{14.4}{\rmdefault}{\mddefault}{\updefault}3\,$Q^2$}}}}}
\put(1898,2520){\makebox(0,0)[lb]{\smash{{{\SetFigFont{12}{14.4}{\rmdefault}{\mddefault}{\updefault}2\,$Q^2$}}}}}
\put(1988,0){\makebox(0,0)[lb]{\smash{{{\SetFigFont{12}{14.4}{\rmdefault}{\mddefault}{\updefault}$Q^2$}}}}}
\put(188,0){\makebox(0,0)[lb]{\smash{{{\SetFigFont{12}{14.4}{\rmdefault}{\mddefault}{\updefault}$Q^2$}}}}}
\put(3248,0){\makebox(0,0)[lb]{\smash{{{\SetFigFont{12}{14.4}{\rmdefault}{\mddefault}{\updefault}3\,$Q^2$}}}}}
\put(3248,2520){\makebox(0,0)[lb]{\smash{{{\SetFigFont{12}{14.4}{\rmdefault}{\mddefault}{\updefault}4\,$Q^2$}}}}}
\put(98,2520){\makebox(0,0)[lb]{\smash{{{\SetFigFont{12}{14.4}{\rmdefault}{\mddefault}{\updefault}2\,$Q^2$}}}}}
\put(5138,0){\makebox(0,0)[lb]{\smash{{{\SetFigFont{12}{14.4}{\rmdefault}{\mddefault}{\updefault}$Q^2$}}}}}
\end{picture}
}
\end{center}
\caption{The ratios of charge squares depend on whether there is 
a brane making more than one non-standard intersection in the 
configuration (which is at most 1) or not (rule (iii)).} 
\end{figure}

\
\

(iv) For each brane its $H_k$ function is given as 
\be
H_k= H^{q_k^2/Q^2}
\ee
where H is defined in (\ref{H}). In the first example given in rule
(iii) we have $H_1=H^4$ and $H_2=H_3=H^3$. On the other hand, in the 
second 
example $H_1=H_2=H^2$ and $H_3=H$.

\
\

(v) The metric is obtained by multiplying the brane and the
transverse directions for $k$'th brane by appropriate powers of
$H_k$ which are 
\be\label{m2h}
\textrm{For i'th SM2-brane:}\begin{cases}
{\textrm{Brane direction}\hs{12}H_i^{-1/3}\cr\cr
 \textrm{Transverse direction}\hs{5}H_i^{1/6}}\end{cases}
\ee
\be\label{m5h}
\textrm{For j'th SM5-brane:}\begin{cases}
{\textrm{Brane direction}\hs{12}H_j^{-1/6}\cr\cr
 \textrm{Transverse direction}\hs{5}H_j^{1/3}}\end{cases}
\ee

\
\

(vi) Upto the $H$-functions, the
overall transverse space takes the form
\be\label{overall}
G_{n,\sigma}^{-\frac{n}{(n-1)}}\left[\,-dt^2\,+\,G_{n,\sigma}\,d
\Sigma_{n,\sigma}^2\,\right]
\ee
where $d\Sigma_{n,\sigma}^2$ is the metric on the unit sphere,
the unit hyperbola or flat space.

\
\

(vii) The 4-form field strength of the configuration is 
\be
F= \sum_{i}F_i + \sum_{j}F_j
\ee
where $F_i$'s are for SM2's given in (\ref{fm2}) and $F_j$'s are for SM5's 
given in (\ref{fm5}).

\
\

With these rules one can construct all possible non-standard intersections 
of SM-branes. Let us mention that it is possible to have intersections 
where 
all intersections are standard, yet the overall intersection
is non-standard. This happens, for example in 
$SM2 \perp\ SM2\perp SM2(-1)$ intersection where branes are located at
$(x_1, x_2, x_3)$, $(x_2, x_4, x_6)$ and $(x_1, x_4, x_5)$ 
with charges $q_1^2=q_2^2=q_3^2=Q^2$. 
We would like to emphasize that there may be more that one configuration 
with a given number of branes, e.g. there are three 
configurations for $SM2 \perp\ SM2\perp
SM2(-1)$. 

\
\

Let us now compare the intersections found here and those standard 
ones obtained in \cite{deger}. To begin with, their construction rules 
are very similar to 
each other. In both of them solutions can be constructed from two 
hyperbolic 
functions $H$ and $G$ defined in (\ref{H}) and (\ref{G}) using rules (v) 
and (vi). The main difference is the number of 
integration constants. In non-standard solutions there are
only
3 parameters, namely $M_1$(or $M_2$), $Q^2$ and $t_0$. In fact, the 
constant $M_1$(or $M_2$) can be removed by a scaling $t\to t/M_1$ followed by 
further scalings of $x_i$ coordinates or by
redefinition of the charge $Q$, if necessary. Therefore, these solutions 
depend only on 2 constants. On the other hand, there are $(3x-1)$ free 
constants in standard intersections where $x$ is the number of branes.
In the standard intersections any brane can 
be removed from the system by certain scalings \cite{deger}. However, in 
our case since branes are indistinguishable from each other this can not 
be done. Another difference is the rule (iv). In the standard 
intersections all 
$H_k$'s are proportional to the square of ${\rm cosh}(t-t_0)$, whereas 
here the power depends 
on charge ratios.
In common with the standard ones, there is always a singularity since 
at 
least 
one of the metric functions vanishes as $t\rightarrow \pm \infty$.
In some of them, like $SM2 \perp SM5(0)$, the singularity is 
milder in the sense that, the curvature scalar (\ref{curvature}) is zero. 
As pointed out in \cite{deger} when the overall tranverse space 
has dimension greater than 2, i.e. ($n=2+m$), then one can flatten 
(\ref{overall}) as
\be\label{sm}
G_{n,\pm}^{-\frac{n}{(n-1)}}\left[\,-dt^2\,+\,G_{n,\pm}\,d
\Sigma_{n,\pm}^2\,\right]\to
dz^2+ 
G_{n-1,\pm}^{-\frac{(n-1)}{(n-2)}}\left[\,
-dt^2\,+\,G_{n-1,\pm}\,d\Sigma_{n-1,\pm}^2\,\right].
\ee
This procedure can be repeated $m$ times. The coordinate $z$
can be used to reduce the solution to type IIA theory. Then, applying S
and T-duality transformations, one can obtain these type of
intersecting S-brane solutions in type IIA and IIB supergravities.

\
\

\subsection{Intersections with non-diagonal terms}

\
\
 
In this subsection we will explore intersections containing
$SM2 \perp SM2(1)$, $SM5 \perp SM5(4)$ and $SM5 \perp SM2(2)$. 
These were excluded in the previous subsection because either there
is a non-diagonal term appearing at the right-hand side of the Einstein's 
equations or $F\wedge F$ is not zero. However, these difficulties can be 
overcome by adding extra branes so that these undesired terms disappear. 
To classify solutions of this type is much harder than those of the 
previous 
section, since there can be more than one non-diagonal direction or there 
can be more than two pairs of branes cancelling the same non-diagonal 
term. We will mainly focus on intersections containing $SM2 \perp SM2(1)$,
and investigate the problem for upto 2 non-diagonal 
directions with only two pairs for every cancellation and comment on what 
happens in other cases. We explicitly checked the validity of 
those rules given in section 2.1 upto 3 
additional branes (but we strongly believe they are valid for others too)
and found that 
they are still respected; there is only a some small modification 
in rule (iii) which we will 
indicate. Hence, it is enough to specify $Q^2$ in (\ref{H}) and the 
relations 
between the charges to describe a solution using rules (v), (vi) and 
(vii).

\
\

Now, let us consider $SM2 \perp SM2(1)$ with SM2's 
located at $(x_1, x_2, x_3)$ and $(x_1, x_2, x_4)$ with charges $(q_1, 
q_2)$ respectively. Then, there is a
non-diagonal term coming from $F_{3ABC}F_{4}^{\, \, ABC}$ in 
equation (\ref{eins1}). Let us add two more 
SM2-branes at $(x_3, x_5, x_6)$ and $(x_4, x_5, x_6)$ with charges $(q_3, 
q_4)$. Now, there is only 1 non-diagonal direction $R_{34}$ for the Ricci 
tensor and if we choose $q_1q_2=-q_3q_4$, then the non-diagonal 
terms cancel each other. Following the method of the previous section, we 
find 
that the spatial 
components of Einstein's equations (\ref{r}) are reduced to two 
integrable equations (\ref{reduce1}) and (\ref{reduce2}) that we 
obtained before provided that $q_1^2=q_4^2$ and $q_2^2=q_3^2$. 
Results given in (\ref{integrability}), (\ref{H}) and (\ref{G}) 
are still valid with $Q^2=(q_1^2+q_2^2)/2$ and $n=4$.
The metric and the 4-form can be obtained using rules (iv)-(vii) and it 
is:
\bea \nonumber
ds^2&=&H^{-1/3}(dx_1^2+...+dx_6^2)+H^{2/3}G_{4,\sigma}^{-4/3}
(\,-dt^2\,+\,G_{4,\sigma}\,d\Sigma_{4,\sigma}^2\,) \\ \nonumber
F&=&q_1*(dx_4dx_5dx_6\Omega_4)+q_2*(dx_3dx_5dx_6\Omega_4)\\ 
&+&q_3*(dx_1dx_2dx_4\Omega_4)+q_4*(dx_1dx_2dx_3\Omega_4).
\label{system1}
\eea
The form of the metric coincides with that of $SM2\perp SM2(-1)$ given in 
(\ref{hodge}). 
Another way of looking at this system is to consider 
the first and the second branes as a single brane with charge 
($q_1^2+q_2^2$). 
(Similarly third and fourth with charge
$q_3^2+q_4^2 =q_1^2+q_2^2$.) Since in this quartet each brane makes one
non-standard intersection (we count the non-diagonal intersections between 
$q_1$ and $q_2$ (or $q_3$ and $q_4$) as standard when applying (rule 
(iii))
, this explains why the charge of a brane making 
only standard intersections should be $Q^2=(q_1^2+q_2^2)/2$ by rule (iii).
If this view point is correct than adding two SM2's at 
$(x_1,x_5,x_7)$ and $(x_1,x_6,x_8)$ should be different than adding SM2's 
at $(x_1,x_5,x_7)$ and $(x_2,x_6,x_8)$. In each case additional branes make 
standard intersections with the brane quartet but among themselves they 
differ. Indeed, in the first one additional branes have charges
$q_5^2=q_6^2=(q_1^2+q_2^2)$/2 whereas in the second they carry 
$q_5^2=q_6^2=q_1^2+q_2^2$ as expected.  

\
\

After looking at many examples we realized that, in addition to the 
restrictions arising from rules (i) and (ii), branes can be 
added to the 
system (\ref{system1}) only when 
they satisfy the following  
exclusion principle:  

\
\

{\it After the addition of 
branes, the 
pair $(q_1, q_2)$ should receive the same number of non-standard 
intersections as that of the pair $(q_3,q_4)$}. 

\
\

For example, we can't add 
just a single SM2 along $(x_1,x_7,x_8)$ to the system (\ref{system1}), 
because it 
makes 2 non-standard 
intersections with the second pair but none with the first. 
This rule can be
understood as follows: To cancel the non-diagonal term we need
$q_1q_2=-q_3q_4$. However, from the rule (iii) we know that
making extra non-standard intersections effects charges, and therefore to
keep the relation valid, we need the above principle.

\
\

Now let us state the modification in rule (iii). {\it When there are 
additional branes, in applying rule (iii) to 
the brane quartet {\rm(\ref{system1})}, only 
external non-standard intersections should be counted.} 
To illustrate, let us add an SM2 at $(x_3,x_7,x_8)$. 
Although it respects the symmetry between the pairs $(q_1,q_2)$ and 
$(q_3,q_4)$ and therefore this addition is possible, it makes non-standard 
intersections only with $(q_2, q_4)$. This system is solvable when
$(q_1^2, q_2^2, q_3^2, q_4^2, q_5^2)=(q_1^2,3q_1^2, q_1^2,3q_1^2,4q_1^2)$ 
which is compatible with rule (iii) if we don't count internal 
non-standard 
intersections of these 4 branes. 
We can add a sixth brane to 
this system at 
$(x_2,x_5,x_7)$ which makes standard intersections only and therefore 
its charge should be $q_6^2=q_1^2$.
One can add at most five SM2-branes to the system (\ref{system1}) and 
there are 15 
different possibilities. For SM5's the maximum number is three and there 
are 3 possibilities. And finally, for combinations we have 7 cases and 
maximum numbers are (2 SM5+ 2 SM2) and (1 SM5 + 3 SM2).

\
\

Intersections with two non-diagonal Ricci components work similarly. 
Let SM2-branes be located at $(x_1,x_2,x_3), (x_1,x_2,x_4), (x_2,x_3,x_5),
(x_2,x_4,x_5)$ with charges $(q_1,q_2,q_3,q_4)$. Now, to cancel $R_{34}$ 
and $R_{15}$ we need $q_1q_2=-q_3q_4$ and $q_1q_3=-q_2q_4$. These imply
$q_1^2=q_4^2$ and $q_2^2=q_3^2$ which also solve Einstein's equations. 
In this solution $Q^2=q_1^2+q_2^2$ and it is given as, 
\bea \nonumber
ds^2&=&H^{-1/3}(dx_1^2+dx_3^2+dx_4^2+dx_5^2)+ H^{-2/3}dx_2^2+
H^{1/3}G_{5,\sigma}^{-5/4}
(\,-dt^2\,+\,G_{5,\sigma}\,d\Sigma_{5,\sigma}^2\,) \\ \nonumber
F&=&q_1*(dx_4dx_5\Omega_5)+q_2*(dx_3dx_5\Omega_5)\\ 
&+&q_3*(dx_1dx_4\Omega_5)+q_4*(dx_1dx_3\Omega_5).
\label{system2}
\eea
This is a tighter system if we want to add more branes, since 
in addition to the above exclusion principle we should also consider
$(q_1,q_3)$ and $(q_2,q_4)$ as pairs. We see that additions are possible 
only when both ($q_1$, $q_4$) and ($q_2$, $q_3$) brane pairs receive the 
same 
amount of non-standard intersections. For example, we can't put a 
single brane at
$(x_1,x_6,x_7)$ since it makes a non-standard intersection with $q_4$ but 
a standard one with $q_1$. 
If we add an SM2 at $(x_6,x_7,x_8)$ which makes non-standard 
intersections with all 
others, then it should carry charge $q_5^2=4(q_1^2+q_2^2)/3=4Q^2/3$. A 
sixth 
brane can be added at 
$(x_1,x_5,x_6)$ which makes only standard intersections with charge 
$q_6^2=(q_1^2+q_2^2)/3$. (If the fifth brane was absent then this brane 
would have charge $(q_1^2+q_2^2)$ because members of the system 
(\ref{system2}) makes only standard intersections between themselves.)
To the system (\ref{system2}) one can add at most 3 SM2's with 7 distinct 
ways, and 3 
SM5's with 3 possibilities. There are 7 cases for their combination with 
maximum numbers (1 SM5+3 SM2) and (2 SM5 + 1 SM2).

\
\

Nothing new happens in intersections containing $SM2 \perp SM5(2)$ or 
$SM5\perp SM5(4)$. Lets 
consider the configuration with branes at 
$(x_1,x_2,x_3), (x_1,..,x_6)$,
$(x_1,x_7,x_8), (x_1,x_4...,x_8)$, with charges $(q_1,..,q_4)$. To get rid 
of the $F\wedge F$ term we need $q_1q_2=-q_3q_4$ which is in accordance 
with Ricci tensor conditions $q_1^2=q_4^2$ and $q_2^2=q_3^2$. We have 
$Q^2=(q_1^2+q_2^2)/2$. Similarly, for $SM5\perp SM5(4)$ pairs located at 
$(x_1,...x_6), (x_1,...x_5,x_7), (x_1,...x_4,x_6,x_8), 
(x_1,...x_4,x_7,x_8)$ with charges $(q_1,...,q_4)$ we find 
conditions $q_1q_2=-q_3q_4$ and $q_1q_3=-q_2q_4$ together with
$q_1^2=q_4^2$ and $q_2^2=q_3^2$ to solve the field equations.
In this solution $Q^2=q_1^2+q_2^2$.

\
\

It is straightforward to construct intersections with more 
number of non-diagonal 
Ricci tensors and we expect them to obey the same principles. 
However, since there are more algebraic conditions to 
satisfy it becomes quite complicated to analyze and after a point  
direct construction becomes easier than trying to decide how to apply 
rules. For example, let us add two SM2's located at  
six branes located at $(x_2,x_3,x_5), (x_1,x_4,x_6)$ with charges 
$(q_5,q_6)$ to the configuration (\ref{system1}).
There are 3 non-diagonal terms along $R_{34}, R_{26}$ and $R_{15}$. 
For cancellation we need $q_1q_2= -q_3q_4$,\, $q_3q_5= -q_2q_6$ 
and $q_1q_5= -q_4q_6$, which are consistent with 
$q_1^2=q_4^2, \, q_2^2=q_3^2$ and $q_5^2=q_6^2$ that are required 
for Einstein's equations. The 
solution is given by $Q^2=(q_1^2+q_2^2+q_5^2)/2$. 
In this case $(q_1,q_2,q_5)$ triplet 
behaves like a single brane. 
As a second case let us consider two 
additional branes to the system (\ref{system1}) at $(x_1,x_2,x_7), 
(x_5,x_6,x_7)$ with charges $(q_5,q_6)$. This time our non-diagonal  
directions are $R_{34}, R_{37}$ and $R_{47}$. Now, cancellations and field 
equations are both satisfied only when $Q^2=q_1^2=....=q_6^2$.

\
\

Another possibility to get rid of unwanted terms is to add more than one 
pair 
of branes. For example, 
consider 6 SM2-branes located at  $(x_1, x_2, x_3)$, $(x_1, x_2, x_4)$,
$(x_3, x_5, x_6)$, $(x_4, x_5, x_6)$, $(x_3, x_7, x_8)$ and $(x_4, x_7, 
x_8)$ with charges $(q_1,....,q_6)$ respectively. Then, to cancel the 
non-diagonal term $R_{34}$ we need $q_1q_2+q_3q_4+q_5q_6=0$. The Ricci 
components are solved as before with conditions 
$q_1^2+ q_2^2= q_3^2+q_4^2= q_5^2+q_6^2$ and 
$q_1^2+q_3^2+q_5^2= q_2^2+q_4^2+q_6^2$. These conditions can be solved 
simultaneously. One consistent choice is $ q_6=q_1=(2+ \sqrt{3})q_2$,
$q_2=q_5$, $q_4=-q_3$ and $q_3^2=2(2+ \sqrt{3}) q_2^2$. In this 
intersection $Q^2=(q_1^2+...+q_6^2)/12$. It 
seems that one can continue adding more branes. However, to 
satisfy all conditions gets harder. 

\
\

Despite the difficulty in classifying 
all the solutions of this section, constructing any one of them is 
straightforward with our method. 
The physical properties of them are 
like those obtained in the previous section. Again none of the branes is 
separable and there are singularities. Once more, they depend only on two 
parameters $(Q^2, t_0)$, but there is a little more freedom for charges 
since usually the proportionality of all charges is not required. Instead, 
some anti-branes are necessary in the system.

\
\

\section{An Intersection with non-vanishing $F\wedge F$ term}

\
\

In this section we would like to investigate an intersection where the 
Chern-Simons term has a non-zero contribution to the 4-form field equation 
(\ref{eins2}).
In all the S-brane solutions constructed until now \cite{s1,s2,s3,deger} 
this term didn't play any role. 
For this purpose, let us consider two SM2-branes located at 
$(x_1, 
x_2, 
x_3)$ and $(x_4, x_5, x_6)$ and an SM5-brane located at $(x_1,..., x_6)$.
Our metric and 4-form field strength are

\be
ds^2=-e^{2A}dt^2\,+\,
e^{2C_1}\,(dx_1^2+dx_2^2+dx_3^2)+\,e^{2C_2}\,(dx_4^2+dx_5^2+dx_6^2)
+\,e^{2D}\,d\Sigma_{4,\sigma}^2
\ee

\be
F= a(t) \, \overline{dt}\wedge \overline{dx_1}\wedge 
\overline{dx_2}\wedge 
\overline{dx_3} + 
b(t) \, \overline{dt}\wedge \overline{dx_4}\wedge 
\overline{dx_5}\wedge
\overline{dx_6} +
q \, d\Omega_4
\ee
where coordinates with bar are in the orthonormal frame. 
The Bianchi identity $dF=0$ is satisfied trivially.
Again we use the gauge condition $A= 3C_1+3C_2+4D$ to simplify the Ricci
tensor. This allows us to choose $A= D + 3g = -C_1-C_2+4g$. We also define

\be
p= a e^{-4C_1-C_2} \, \, , \hs{4} r= b e^{-4C_2-C_1}
\ee
Note that 
$F\wedge F$ is not zero and from the 4-form field equation (\ref{eins2}) 
we get two constraints:

\bea
p'&=& q\, r\, e^{6C_2}\\
r'&=& - q\, p\, e^{6C_1}
\eea

Spatial components of the Ricci tensor gives

\bea
C_1'' &=& -\frac{1}{3}p^2\, e^{6 C_1} + \frac{1}{6} r^2 \, e^{6C_2} - 
\frac{q^2}{6} \, e^{6C_1+6C_2}\\
C_2'' &=& \frac{1}{6}p^2\, e^{6 C_1} - \frac{1}{3} r^2 \, e^{6C_2} - 
\frac{q^2}{6} \, e^{6C_1+6C_2}\\
g''&=& -3\, \sigma \, e^{6g}
\eea
Notice that when $q$ is zero this system of equations reduces to 
equations (\ref{-1})-(\ref{-3})
which we wrote in the previous section for $SM2 \perp SM2(-1)$ 
with $p=q_1$ and $r=q_2$. 

\
\

The last equation is 
decoupled from others and its solution is given in (\ref{G}). However, the 
remaining 
two are non-linear equations and we couldn't solve it analytically. But, 
numerical techniques can be used to see the behaviour of the metric 
functions. The time component of the Ricci tensor is necessary for a 
consistent choice of the initial values and that gives,

\be
-12(C_1')^2 -12(C_2')^2 - 12C_1'C_2'+24 M_2^2 = p^2e^{6C_1} + r^2 e^{6C_2}
+ q^2e^{6C_1+6C_2}
\ee
where $M_2^2$ comes from the integrability of the function $g$ 
given in (\ref{g2}).
In picking up initial values it is also useful to notice $2q(C_1'' - 
C_2'')= 
(pr)'$. Actually there are several options for initial values which 
change the asymptotic values of the functions $p$ and $r$. 
However, 
we observed that the asymptotic behaviour of the metric 
functions $e^{2C_1}$ and $e^{2C_2}$ do not depend much on these choices 
and 
they always approach to zero as $t\rightarrow \infty$ which signals a 
singularity (figure 2).

\begin{figure}
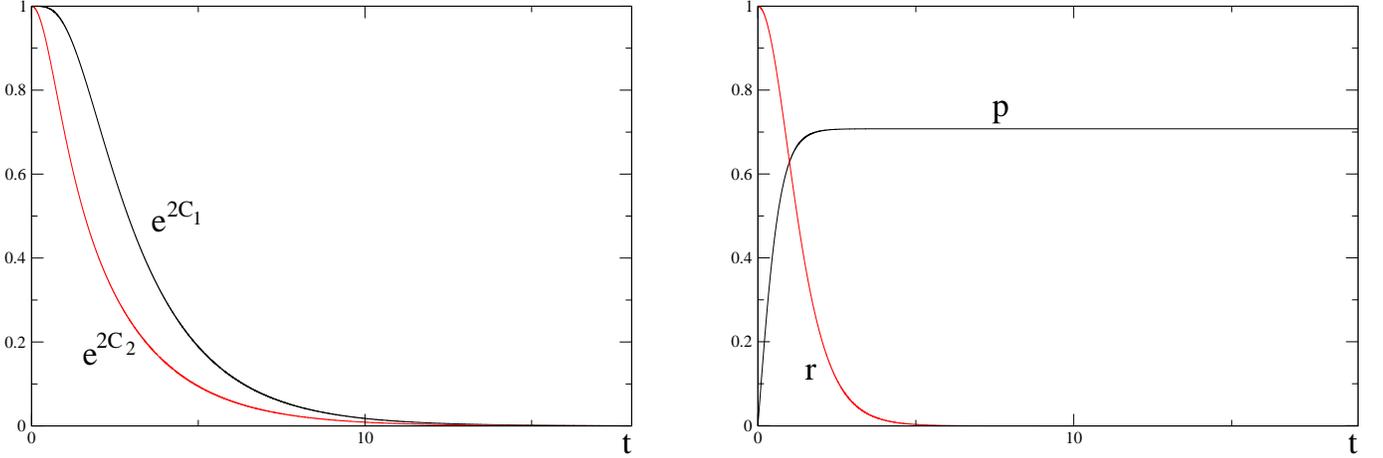

\centerline{\epsfxsize=3.3truein
\epsffile{metric.eps}
\hspace{0.4in}
\epsfxsize=3.3truein
\epsffile{form.eps}
}
\caption{The metric and form functions with initial values 
p(0)= $C_1(0)=C_2(0)= C_1'(0)=C_2'(0)$=0, r(0)=1 and q=1, $M_2^2=1/12$.}
\end{figure}

\
\

\section{Conclusions}

\
\

\hs{5}In this paper, we have constructed a large number of 
orthogonally intersecting S-brane solutions in $D=11$ supergravity 
which do not have  
supersymmetric analogs. They depend only on 
2-parameters and there is no parameter to distinguish one brane from 
the 
other. It is remarkable that all of them can be obtained just by using two 
functions 
defined in (\ref{H}) and (\ref{G}) and a simple set 
of axioms which are the main results of this paper.
It is nice to see that despite the absence of supersymmetry 
both the solutions obtained in this paper and those standard 
intersections obtained before in \cite{deger} have similar construction 
rules which is a  beautiful reflection of the structure of the
$D=11$ supergravity field equations.

\
\

We also analyzed a solution where the Chern-Simons term has a non-trivial
effect to the field equations. 
Unfortunately, as in other  
supergravity S-brane solutions  
there is a naked singularity. Recently there has been some progress in 
resolution of these singularities \cite{rec2, rec6, rec8} by using 
gravitational backreaction of tachyon matter. 

\
\

It seems that there is a close relationship between S-brane solutions and  
non-extremal p-brane solutions. The technique to obtain S-branes was used 
to construct black p-branes in \cite{s2} and the similarity of the solutions 
are obvious. It is known that upto Wick rotations one can sometimes map a 
cosmological solution into a p-brane solution (see for example 
\cite{map1,pope2}). It would be nice to understand this connection for 
S-branes.
Then, one can see what the S-brane analogs of monopole and wave 
solutions of the $D=11$ supergravity are and study their intersections. 
Also it would be interesting to find out the static version of the 
$SM2 \perp SM2(-1)$ configuration. We expect that most of the solutions
we obtained have static counterparts. For instance, 
non-supersymmetric but extremal $M2 \perp M5(0)$ 
solution is discussed in \cite{argurio}. In addition, several 
non-extreme static brane 
configurations where charges and harmonic functions are not independent 
are found in \cite{ohta2, ohta3}. In these the common tangent 
directions are not Poincar\'{e} invariant.

\
\

Certainly more work is required in order to understand how to 
distinguish standard and non-standard intersections from 
each other. M-brane intersections can be divided into 3 classes
by looking at their binding energies (see \cite{pope3} for a review). 
A generalization of ADM mass for S-brane backgrounds is necessary to do 
such comparison. We hope that our solutions will be useful in this 
respect.

\
\

There are various 
possible generalizations of 
the solutions we found. 
First of all, instead of adding anti-pairs 
to obtain intersections that are worked out in section 2.2, 
one may seek a different
ansatz for the 4-form field strength. 
Secondly, all our solutions are orthogonal and one may consider 
intersections with angles. Thirdly, one may assume curved 
world-volume for S-branes and study their intersections. Of course, when 
both the 
world-volume and the 
transverse space are curved equations may not be solvable in general
\cite{pope2}. And finally, our solutions depend only on time 
and one may try to find localized ones where solutions depend on other 
transverse directions as well. 

\
\

\subsection*{Acknowledgements}
\hs{4} We would like to thank A. Kaya for stimulating  discussions and 
T. Rador for computer help.

\
\


\begin{thebibliography}{25}

\bibitem{polc}J. Polchinski, {\sl Dirichlet-Branes and Ramond-Ramond
Charges}, Phys. Rev. Lett. 75 (1995) 4724, hep-th/9510017.

\bibitem{s1}M. Gutperle and A. Strominger, {\sl Spacelike Branes}, JHEP 
0204 (2002) 018, hep-th/0202210.

\bibitem{s2}C.M. Chen, D.M. Gal'tsov and M. Gutperle, {\sl S-brane 
Solutions in Supergravity Theories}, Phys.Rev. D66 (2002) 
024043, hep-th/0204071.

\bibitem{s3}M. Kruczenski, R.C. Myers and A.W. Peet, {\sl Supergravity 
S-Branes}, JHEP 0205 (2002) 039, hep-th/0204144.

\bibitem{s4}S. Roy, {\sl On Supergravity Solutions of Space-like 
Dp-branes}, JHEP 0208 (2002) 025, hep-th/0205198.

\bibitem{deger} N.S. Deger and A. Kaya, {\sl Intersecting S-Brane 
Solutions of D=11 Supergravity}, JHEP 0207 (2002) 038, hep-th/0206057.

\bibitem{rec1} J.E. Wang, {\sl Spacelike and Time Dependent Branes from 
DBI}, JHEP 0210 (2002) 037, hep-th/0207089.

\bibitem{sen} A. Sen, {\sl Time Evolution in Open String Theory}, JHEP 
0210 (2002) 003,  hep-th/0207105.

\bibitem{rec2} A.Buchel, P. Langfelder and J. Walcher, {\sl Does the 
Tachyon Matter?}, Annals Phys. 302 (2002) 78, hep-th/0207235.

\bibitem{toda3} V.D.Ivashchuk , {\sl Composite S-brane solutions related
to Toda-type systems}, Class. Quant. Grav.20 (2003) 261, hep-th/0208101.

\bibitem{rec3} A. Strominger, {\sl Open String Creation by S-Branes}, 
hep-th/0209090.

\bibitem{rec5} K. Hashimoto, P.M Ho and J.E. Wang, {\sl S-brane Actions}, 
hep-th/0211090.

\bibitem{rec6}  A. Buchel and J. Walcher, {\sl The Tachyon does Matter}, 
hep-th/0212150.

\bibitem{ohta} N. Ohta, {\sl Intersection Rules for S-Branes}, 
hep-th/0301095.

\bibitem{zavala}  C.P. Burgess, P. Martineau, F. Quevedo, G. Tasinato 
and I. Zavala, {\sl Instabilities and Particle Production in S-Brane 
Geometries}, hep-th/0301122.

\bibitem{rec7} A. Maloney, A. Strominger and X. Yin, {\sl S-Brane 
Thermodynamics},  hep-th/0302146.

\bibitem{rec8} F. Leblond and A.W. Peet, {\sl SD-brane gravity fields and 
rolling tachyons}, hep-th/0303035.

\bibitem{rec9} K. Hashimoto, P.M. Ho, S. Nagaoka and J.E. Wang, {\sl Time 
Evolution via S-branes}, hep-th/0303172.

\bibitem{strominger} A. Strominger, {\sl The dS/CFT Correspondence},
JHEP 0110 (2001) 034, hep-th/0106113.

\bibitem{behrndt} K. Behrndt and S. Foerste, {\sl String--Kaluza--Klein 
Cosmology}, Nucl.Phys. B430 (1994) 441, hep-th/9403179.

\bibitem{pope1}  H. Lu, S. Mukherji, C.N. Pope and K.W. Xu, {\sl 
Cosmological Solutions in String Theories}, Phys.Rev. D55 (1997) 7926,
hep-th/9610107.

\bibitem{pope2}  H. Lu, S. Mukherji and C.N. Pope, {\sl From p-branes to 
Cosmology}, Int.J.Mod.Phys. A14 (1999) 4121, hep-th/9612224.

\bibitem{ovrut1} A. Lukas, B.A. Ovrut, D. Waldram, {\sl Cosmological 
Solutions of Type II String Theory}, Phys.Lett. B393 (1997) 65, 
hep-th/9608195.

\bibitem{ovrut2} A. Lukas, B.A. Ovrut, D. Waldram, {\sl String and 
M-Theory Cosmological Solutions with Ramond Forms}, Nucl.Phys. B495 (1997) 
365, hep-th/9610238.

\bibitem{toda2} V.D. Ivashchuk and V.N. Melnikov, {\sl Multidimensional 
Classical and Quantum Cosmology with Intersecting p-branes}, J.Math.Phys. 
39 (1998) 2866, hep-th/9708157. 

\bibitem{toda4} V.D. Ivashchuk and V.N. Melnikov, {\sl Exact solutions in 
multidimensional gravity with antisymmetric forms}, Class.Quant.Grav. 18 
(2001) R82, hep-th/0110274.

\bibitem{har1}A.A. Tseytlin, {\sl Harmonic superpositions of
M-branes}, Nucl. Phys. B475 (1996) 149, hep-th/9604035.

\bibitem{har2}K. Behrndt, E. Bergshoeff, B. Janssen, {\sl Intersecting 
D-Branes in ten and six dimensions}, Phys.Rev. D55 (1997) 3785, 
hep-th/9604168.

\bibitem{har3}J.P. Gauntlett, D.A. Kastor and  J.Traschen, {\sl
Overlapping Branes in M-Theory}, Nucl. Phys. B478 (1996) 544, hep-th/9604179.

\bibitem{har4} E. Bergshoeff, M. de Roo, E. Eyras, B. Janssen, J.P. van 
der Schaar, {\sl Multiple Intersections of D-branes and 
M-branes}, Nucl.Phys. B494 (1997) 119, hep-th/9612095.

\bibitem{map1} F. Larsen and F. Wilczek, {\sl Resolution of Cosmological 
Singularities}, Phys.Rev. D55 (1997) 4591,  hep-th/9610252.

\bibitem{argurio} R. Argurio, {\sl Brane Physics in M-theory}, 
hep-th/9807171. 

\bibitem{ohta2} N. Ohta and T. Shimizu, {\sl Non-Extreme 
Black Holes from Intersecting M-Branes}, 
Int. J. Mod. Phys. A13 (1998) 1305, hep-th/9701095.

\bibitem{ohta3} N. Ohta, {\sl Intersection Rules for Non-Extreme 
$p$-Branes}, Phys.Lett. B403 (1997) 218,  hep-th/9702164.

\bibitem{pope3} H. Lu, C.N. Pope, T.R. Tran and K.W. Xu, {\sl 
Classification of p-branes, NUTs, Waves and Intersections}, Nucl.Phys. 
B511 (1998) 98, hep-th/9708055. 

\end{thebibliography}
\end{document}